\documentclass{aastex}
\usepackage{emulateapj5}
\begin{document}

\title{ Optical Observations of the Black Hole Candidate
XTE J1550-564 During Re-Flare and Quiescence}

\author{Raj K. Jain\altaffilmark{1,2}, Charles D. Bailyn\altaffilmark{2}, 
Jerome A. Orosz\altaffilmark{3}, Jeffrey E. McClintock\altaffilmark{4},
Gregory J. Sobczak\altaffilmark{4}, and Ronald A. Remillard\altaffilmark{5}}

\altaffiltext{1}{Department of Physics, Yale University,
 P.\ O.\ Box 208120, New Haven, CT 06520-8120, raj.jain@yale.edu} 
\altaffiltext{2}{Department of Astronomy, Yale
 University, P.\ O.\ Box 208101, New Haven, CT 06520-8101,
 bailyn@astro.yale.edu}
\altaffiltext{3}{Sterrenkundig Instituut, Universiteit Utrecht, 
Postbus 80.000, 3508 TA Utrecht, The Netherlands, 
 J.A.Orosz@astro.uu.nl}
\altaffiltext{4}{Harvard-Smithsonian 
Center for Astrophysics, 60 Garden Street, Cambridge, MA 02138-1516, 
jem@cfa.harvard.edu, gsobczak@head-cfa.harvard.edu}
\altaffiltext{5}{Center for Space Research,
Massachusetts Institute of Technology, Cambridge, MA 02139-4307, 
rr@space.mit.edu}

\begin{abstract}
We report optical monitoring of the soft X-ray transient XTE J1550-564 during
the 1999 season (4 January 1999 to 24 August 1999).  The first optical
observations available in 1999 show that the peak ``re-flare'' brightness had
exceeded the peak brightness of the initial optical flare in September 1998 by
over half a magnitude.  We compare the optical re-flare light curves with the
total X-ray flux, the power-law flux and disk flux light curves constructed
from the spectral fits to RXTE/PCA data made by Sobczak et al.\ (1999, 2000).
During the first 60 days of the observed optical re-flare, we find no
correspondence between the thermal component of the X-rays often associated
with a disk and the optical flux -- the former remains essentially flat
whereas the latter declines exponentially and exhibits three substantial dips.
However, the power law flux is anti-correlated with the optical dips,
suggesting that the optical flux may by up-scattered into the X-ray by the hot
corona.  Periodic modulations were discovered during the final stage of the
outburst (May to June), with P=$1.546\pm0.038$ days, and during quiescence
(July and August), with P=$1.540\pm0.008$ days.  The analysis of the combined
data set reveals a strong signal for a unique period at P=$1.541\pm0.009$
days, which we believe to be the orbital period.
\end{abstract}


\section{Introduction}

X-ray binaries, which contain a neutron star or a black hole, are among the
brightest and most carefully studied X-ray sources.  A subclass of X-ray
binaries, called soft X-ray transients (SXTs) or X-ray novae, are mass
transferring binaries in which the Roche filling mass donor is a dwarf or
sub-giant and the accreting compact object is a black hole or a neutron star.
SXTs undergo long periods of quiescence (when the X-ray luminosity is $\le
10^{33}$ergs s$^{-1}$) which is occasionally interrupted by luminous X-ray and
optical outbursts lasting anywhere from $\sim 10$ to $\gtrsim$ 100 days
(Tanaka \& Shibazaki 1996; Chen, Shrader, \& Livio 1997).  The mass function,
which is a firm lower limit on the mass of the compact object, can be obtained
from optical observations for systems which are bright in quiescence and
dominated by the secondary star.  The inclination and mass ratio can also be
determined, leading to a complete knowledge of the orbital parameters and the
masses.

Nine SXTs have been shown to contain a black hole (van Paradijs \& McClintock
1995; Bailyn et al.\ 1998; Orosz et al.\ 1998a, Filippenko et al. 1999), as
their mass functions, or the constrained masses, exceed the maximum stable
limit of a neutron star, which is $\approx 3M_{\sun}$ (Chitre \& Hartle 1976).
In comparison to the numerous observations of SXTs during quiescence (Bradt \&
McClintock 1983; van Paradijs \& McClintock 1995), thorough optical
observations of SXTs during outburst are rare.  Obtaining good outburst light
curves in the optical and in X-ray bands requires an all sky monitor on board
an X-ray satellite and uninterrupted access, throughout the year, to an
optical telescope.  There have been several all sky monitors on previous X-ray
satellites, but the opportunity to obtain excellent X-ray light curves of
transients dramatically increased with the introduction of the {\em Rossi
X-ray Timing Explorer} (RXTE).  Generally, SXT outbursts are covered with a
series of pointed RXTE observations every $\sim$ few days (with energy range
between 2-200 keV and time
resolution as low as 1$\mu$s), while the All Sky Monitor (ASM; Levine et
al. 1996) provides more frequent and less sensitive coverage by scanning most
of the X-ray sky every few hours.  On the other hand, comparable optical light
curves have been difficult to obtain, due to inflexible telescope scheduling,
unpredictable weather conditions, and the limitations posed by the position of
the object in the night sky.  To circumvent the difficulties posed by
conventional telescope scheduling, we have used the Yale 1-m telescope located
in CTIO and operated by the YALO (Yale, AURA, Lisbon, and Ohio State)
consortium.  Observations are requested, nightly, via the web and taken by two
permanent staff observers, providing the flexible scheduling necessary for
studying new outbursts.

Simultaneous X-ray and optical data on SXT outbursts are important for several
reasons.  First, optical observations provide diagnostics of the properties of
the outer accretion disk.  The optical emission of a black hole SXT is
generally thought to arise in the outer accretion disk due to reprocessing of
X-rays that are emitted by the inner disk (van Paradijs \& McClintock 1995;
King \& Ritter 1998).  Recent work suggests that the outer disk must be
significantly warped in order to explain the high intensity of the reprocessed
flux (Esin et al. 1999; hereafter EKMN).  A warped disk was also invoked by
Esin, Lasota and Hynes (hereafter ELH) to explain the different decay profiles
of both optical and X-ray light curves of GRO 1655-40, however the optical
light curve was sparsely sampled.  In another model for the accretion
geometry, the inner accretion disk is replaced by an optically thin ADAF
(Narayan, Mahadevan, \& Quataert 1998, and references therein).  In this case,
optical data provide important information required to constrain the spectra
predicted by this composite model.

Second, the features of the optical and X-ray light curves are often
correlated, with delays spanning several seconds to days.  Such delays are
related to interesting physical timescales, such as the viscous and thermal
timescales or the light-crossing time, which in turn provide information about
the $\alpha$-viscosity parameter, radial extent of the disk, and accretion
rates.  For example, strong evidence in support of the composite ADAF and thin
disk model for quiescent SXTs (Lasota 1999) comes from a recent observation of
a 6 day delay between the rise of the 1996 optical and X-ray outbursts of GRO
1655-40 (Orosz et al.\ 1997), which has been successfully modeled by an
accretion flow consisting of an outer thin disk and an inner ADAF region
(Hameury et al.\ 1997).  However, it is sometimes the case that the X-ray and
optical light curves are uncorrelated (e.g.\ XTE J1550-564, Jain et al.\ 1999a;
this paper).  Such contrasting X-ray and optical behavior poses a challenge to
models which simply predict an exponentially declining light curve at all
wavelengths and to models which simply assume optical photons come from the
X-ray heated outer disk.

The soft X-ray transient XTE J1550-564 was discovered with the ASM on the RXTE
on September 6, 1998 (Smith et al.\ 1998).  The optical counterpart (Orosz,
Bailyn, \& Jain 1998b; Jain et al.\ 1999a) as well as the radio counterpart
(Campbell-Wilson et al.\ 1998) were identified shortly thereafter.  Spectra of
the optical counterpart during outburst show prominent H$\alpha$ emission as
well as broad and weaker emission lines from H$\beta$ and He II
(S\'{a}nchez-Fern\'{a}ndez et al.\ 1999, hereafter S-F).  Based on the
equivalent widths of Na D absorption lines and the diffuse interstellar
features, S-F find $E(B-V)=0.60$ and infer a distance of 2.5 kpc.  The object
continued to increase in brightness, reaching a maximum of $6.8$ crab during a
bright flare, making XTE J1550-564 the second brightest new X-ray source
detected by RXTE (Sobczak et al.\ 1999).

We have published the X-ray spectral, X-ray timing, and optical results for the
period 1998 September 6 to October 28: Sobczak et al.\ 1999 (Paper I);
Remillard et al.\ 1999 (Paper II); and Jain et al.\ 1999a (Paper III).  We
observed the canonical spectral states of a black hole SXT, both the high and
low frequency QPOs, and optical responses to large amplitude X-ray
fluctuations.

As shown by the ASM X-ray light curve in Fig.~\ref{fig1}d, near the end of
1998, the source appeared to be heading towards quiescence (Paper I).
However, a prolonged re-flare was observed in the X-ray and optical between
MJD 51162 (1998 December 15) and $\sim$ MJD 51350 (1999 June 20). In this
paper we report our optical observations and results for this 27 week
period. We present comparisons between the optical and X-ray light curves as
well as evidence for periodic behavior in the optical light curve both during
the final stages of outburst and quiescence. The spectral and timing analysis
of the RXTE proportional counter array (PCA) data as well as their
correlations for both the 1998 and the 1999 data, up to the end of RXTE Gain
Epoch 3 (i.e. before 16:30 (UT) on 1999 March 22) can be found in Paper I, II
and Sobczak et al.\ 2000 (hereafter Paper IV).

In \S 2 we discuss the observation and data reduction methods.  In \S 3 we
present the re-flare light curve and discuss some of its salient features. To
facilitate the discussion we divide the re-flare light curves into four
intervals. In \S 4 we describe evidence for periodic behavior of the re-flare
light curve shortly before and during quiescence. We also discuss the possible
origin of the periodic behavior.  In \S 5, we compare the $B$,V and I light
curves with the X-ray light curve. The total X-ray flux can be decomposed into
contributions from a disk blackbody component and a power law component (Paper
IV) and we compare the behavior of each of these components with the optical
light curve.  In \S 6 we provide a discussion which compares the present data
set with existing models and in \S7 we end with a brief summary.


\section{Observations and Reductions} 

We obtained standard (Johnson) $B$, $V$, and (Kron) $I$ and wide bandpass
Johnson $R$ photometry using the Yale 1m telescope at CTIO, which is currently
operated by the YALO consortium (Bailyn et al. 2000).  Data were acquired
using the ANDICAM optical/IR camera which contains a TEK $2048\times 2048$ CCD
with $10.2 \times 10.2$ arcmin$^2$ field of view with a scale of 0.3 arcsec
pixel$^{-1}$.  There are gaps in the light curve in April due to equipment
failure caused by a lightening strike, and early July when the IR array was
brought into operation.  No IR data was obtained for XTE J1550-564.

Paper III reported data between 1998 September 8.99 and October 26.9.  Due to
observational constraints imposed by the location of the object in the night
sky, we were not able to re-embark on our $BVI$ observing campaign until 1999
January 4.  After this, at least one exposure was taken per filter for each
night the weather and equipment permitted.  We concluded the $B$, $V$, and $I$
observations on 1999 May 7.3 as the source brightness approached the limiting
magnitudes.  During 1999, we obtained 241, 280, and 262 exposures in $B$, $V$,
and $I$, respectively, with observation times ranging between 300 and 1200
seconds (see Table~\ref{tab:journal} for journal of observations).

Although the object was too faint for the $B$, $V$, and $I$ filters by
mid-May, we were able to continue observing using the Johnson $R$ (or wide
$R$) filter. The bandpass of this filter is approximately the same as the
combined bandpass of the standard Kron $R$ and $I$ filters, which makes the
wide $R$ filter ideally suited for studying faint red objects. Unfortunately,
calibration standards, e.g. those from Landolt, are not available in Johnson
$R$ (hereafter simply $R$), therefore we will work exclusively with
differential magnitudes.  Observations in $R$ started on 1999 April 27 and
ended on 1999 August 24. We obtained 171 images with exposure times ranging
from 900 to 1200 seconds (see Table~\ref{tab:journal}). The last 82 exposures
consist of 41 pairs of 2 consecutive 1200 second integrations, which were
summed to increase the signal to noise ratio.  The seeing throughout the $B$,
$V$, $R$, and $I$ campaign varied from 1.2'' to 2.4'' with a median value of
1.5''.  Ninety percent of the exposures had seeing less than 1.8''.

Preliminary data reduction to make the bias, overscan and flat field
corrections, were handled using customized scripts which invoke IRAF routines.
The light curve was obtained using IRAF versions of DAOPHOT and the
stand-alone code DAOMATCH and DAOMASTER (Stetson 1987, 1992a, 1992b; Stetson,
Davis, \& Crabtree 1990).  The DAOPHOT instrumental magnitudes in $BVI$ were
calibrated to standard scales using 4 neighbor stars, whose calibrated
magnitudes were obtained previously (see Fig.~\ref{fig0}, Table~\ref{tab:mag}
and Paper III)\footnote{S\'{a}nchez-Fern\'{a}ndez et al.  have also calibrated
9 neighbor stars in the XTE J1550-564 field.  To check for consistency, we
used our local standards to calibrate the YALO instrumental magnitudes
corresponding to the local standards listed in Table 1 of S-F.  The average
deviation between their calibrated magnitudes and the values we obtained were
0.03, 0.07, and 0.05 in $B$, $V$, and $I$, respectively. However, in computing
this average we excluded stars 4, 5 and 8 as the magnitudes in S-F
contain typographical errors (S\'{a}nchez-Fern\'{a}ndez 2000).}  


\section{The Re-flare Light Curve}

Based on the ASM light curve from late October and November of 1998,
XTE~J1550-564 appeared to be heading steadily towards quiescence.  However, by
December 15th, the X-ray flux of the source had clearly increased, as the ASM
count rate from this time is double the average count rate of the previous 35
days (see Fig.~\ref{fig1}d).  This was the beginning of a X-ray re-flare which
eventually lasted for $\sim 140$ days.  The optical monitoring did not cover
the earliest stages of the re-flare, and by the time we resumed observations
in early January, the source had brightened by 0.54, 0.8, and 0.60 magnitudes
in $B$, $V$, and $I$, respectively, compared to the values obtained during the
optical flare in September 1998 (see Fig~\ref{fig1}, Paper III).

The general features of the light curve of the 1999 re-flare data differ
substantially from the 1998 light curve (Paper III).  With the exception of
one large flare, the 1998 optical light curves can be described as an
exponential decay of the flux with constant e-folding timescale of $\sim 30$
days.\footnote{An exponential decline of the flux gives rise to a linearly
decaying optical light curve -- since magnitudes are logarithms of the flux.
When we say the optical light curve is exponentially declining, we are
referring to the flux and not the magnitude.}  Low amplitude flares, dips, and
periodic signals in the light curve were generally not observed.  Slight
deviations from the smooth exponential decline can be found during the first
ten and the last seven days of the 1998 light curve, which can better be
described as plateau periods.  On the other hand, the re-flare light curve
exhibited a wider range of phenomena.  In Fig.~\ref{fig2} we have plotted the
$V$, $V-I$, $R$, ASM count rate and ASM hardness ratio (HR2) light curves
(where HR2 is defined as the ratio of 5-12 counts and 3-5 keV counts). The $R$
light curve was placed on the same scale as the $V$ light curve by adding an
offset determined from the data taken in late April when both filters were
used.  This was done only to facilitate the comparison of the $R$ light curve
with other colors, and these magnitudes should not be taken to represent true
$R$ magnitudes since $V-R$, presumably, is not zero.  To draw attention to the
salient features of the light curve, we have divided the observation run into
four intervals (with days in units of MJD): Interval~1 (51180 to 51229),
Interval~2 (51230 to 51313), Interval~3 (51314 to 51345), and Interval~4
(51346 to 51420) -- see Table~\ref{tab:int} \& Fig.~\ref{fig2}.


\subsection{Interval 1 -- Large Flux and Slow Decay}

During Interval~1 the optical brightness decreased very slowly, and roughly
exponentially at a rate of $\sim$0.009 mag/day in $V$, corresponding to an
e-folding time of 121 days (see Figs.~\ref{fig1},~\ref{fig2}, and~\ref{fig3}).
During this time, there were two sharp dips in the optical light curve, which
occurred around MJD 51200 and 51215 with amplitudes of $\sim 0.33$ and 0.25 in
$B$, $\sim$0.32 and 0.25 in $V$, and $\sim 0.32$ and 0.24 in $I$ (see
Figs.~\ref{fig2} and \ref{fig3}).  Each of the dips lasted $ \sim 3$ days.
Other than these two features, the light curve does not deviate significantly
from an exponential decay.  The $V-I$ color increased slowly at a rate of
about $6\times10^{-4}$ magnitudes/day, which corresponds to a slight reddening
of the system.  We do not find periodic signatures from the de-trended data
from Interval 1.


\subsection{Interval 2 -- Steep Decline with Random Variability}

Around MJD 51231, the optical flux of XTE~J1550-564 began to drop dramatically
and deviate strongly from the smooth exponential decline found during Interval
1 (see Figs.~\ref{fig2} \& \ref{fig4}).  Within the first three days of
Interval 2, between MJD 51230.34 and 51233.40, the $V$-band flux dropped by
$\sim$0.6 magnitudes, nearly 22 times more rapidly than the rate found during
Interval 1. Subsequently, the source brightened by $\sim 0.21$ mag in the next
5 days (or by MJD 51238.27) and then decreased again in brightness by $\sim
0.85$ magnitudes during the next 4 days.  Between MJD 51244 and 51300 the
optical light curve is much less variable and resembles an exponential decline
with an e-folding time of $\sim 27$ days, or a decay of $\sim 0.04$ mag/day,
which is $\sim 5$ times larger than the value found during Interval 1. There
is also flickering in the light curve, of order $0.1\sim0.2$ magnitudes
superposed on this decaying component.  As mentioned in section II, there is a
substantial gap in the light curves between MJD 51267 and 51294; however it is
likely that the $V$ light curve during the gap continued with the exponential
decline, as the data immediately after the gap are still consistent with the
exponential decay. Finally, the $V$ light curve tapers into a brief plateau as
the rate of decay between MJD 51300 and 51313 diminishes to $\sim 1.4\times
10^{-3}$ mag/day, corresponding to an e-folding time of $\sim 776$ days.

We began our $R$-band observations on MJD 51295.18 (see Fig.~\ref{fig2}).  The
$R$-band data is essentially flat between MJD 51295 and 51312 with minimal
decay of only $\sim 3.1\times 10^{-3}$ mags/day.  It is difficult to compare
the $R$ light curve to the $V$ light curve due to the small number of data
points which overlap in time.  However, the transition from an exponential
decay to a plateau which is seen in the $V$ light curve during the last few
days of Interval 2, just before MJD $\sim 51300$, is difficult to see in the
$R$ light curve (Fig.~\ref{fig2} \& \ref{fig4}).  It may be that this
transition occurred earlier in $R$, prior to our first observations in this
filter.


\subsection{Interval 3 -- Exponential Decay with Periodic Variability}
During Interval 3 (between MJD 51313 and 51345) and the next interval, data
were collected only in $R$ as the object was too faint for the narrower $V$
and $I$ filters.  Contrary to the indications near the end of Interval 2 that
the source might be approaching quiescence, its brightness continued to
decline throughout Interval 3 (see Figs.~\ref{fig2} and \ref{fig5}).  We find
no dips of the sort found in Intervals 1 and 2.  The light curve can be
described as a superposition of a sinusoidal component and an exponential
decay, with a rate of $\sim 0.048$ mags/day or e-folding time of $\sim 22.6$
days. This value is comparable to the e-folding time of $\sim 27$ days found
during the exponentially decaying phase of Interval 2.  Detailed discussion of
the periodic component of the light curve during this interval follows in
section 4.  

\subsection{Interval 4 -- Quiescence}

Between MJD 51346 and the end of the observing campaign on MJD 51420, we
believe the source was in or close to full quiescence.  A linear fit to the
light curve reveals a slight decreasing trend in the brightness, with a rate
of $\sim 3.0\times 10^{-3}$ mag/day or an e-folding time of 362 days (see
Fig.~\ref{fig2} \&\ref{fig6}).  As there was still a decaying component to the
light curve, there remains the possibility that true quiescence had not been
reached.  During Interval 4, the source brightness was approaching the
limiting magnitude of the instrument, hence we averaged two consecutive 1200s
integrations to increase the signal to noise ratio.  In addition, due to
inclement weather and engineering tests, we were unable to get data on a daily
basis, with notable gaps between MJD 51368 and 51378 and MJD 51395 and 51406,
thus limiting our analysis.

It is important to establish the quiescent magnitude corresponding to a
standard set of filters to facilitate comparisons with other sources.
Unfortunately the quiescent data were taken using the wide-$R$ filter, which
is difficult to calibrate using conventional methods (e.g.\ via Landolt
standards).  However, during the last ten days of Interval 2, the light curve
is relatively flat, and moreover we have data available in all four filters.
Assuming the source did not change color significantly, we can estimate the
$B$, $V$, and $I$ magnitudes in quiescence by adding the difference between
the average $R$ band magnitudes from the end of Interval 2 and 4 to the $B$,
$V$, and $I$ magnitudes from the end of Interval 2.  Clearly, the assumption
of constant color may be wrong as the $V-I$ color changed substantially,
during the steep decline of Interval 2 (see Figure~\ref{fig2}).  If we adopt
the change in $V-I$ color observed during Interval 2, $\sim 0.34$ mags as an
additional source of uncertainty, we obtain the mean quiescent $V$ and $I$
magnitudes of $21.3\pm 0.4$ and $19.01\pm0.4$, respectively. This value is
similar to that of XN~Oph 1977 which has a quiescent $m_{V}=21.3$ (Remillard
et al. 1996).  There are only four data points in the
$B$-band data during the last ten days of Interval 2 and the associated errors
are large due to the faintness of the source. Hence the quiescent $B$ value of
$22.7\pm0.4$ must be taken with some caution.  However, this value is
consistent with the quiescent magnitude of $22\pm 0.5$ which was obtained from
aperture photometry of the digitized SERC J plate No. J2977 (Paper III).


\section{Timing Analysis of Intervals 3 and 4 }

We find strong evidence for periodic signatures from Intervals 3 (Jain et
al. 1999b) and 4.  Before applying the period searching algorithms, the data
had to be de-trended in order to minimize the contribution to the power
spectrum from the decaying component of the light curve. The decaying
component is subtracted by a linear fit for the data from Interval 3.  During
interval 4 there is a large gap in the data which roughly separates the
light curves into two shorter segments.  We took account of the decaying
component during Interval 4 by subtracting a linear fit to the entire Interval
4 data set as well as by scaling the average value of the data from both
segments to a common value.   Subsequently, we used the publicly available time
series analysis package PERIOD, which is distributed by project Starlink, to
search for periodic signatures in the time series data.  Among the various
methods for detecting periodic signatures from time series data, we used three
independent methods: the CLEAN algorithm (Roberts, Leh\'ar, \& Dreher 1987),
the Lomb-Scargle periodogram (L-S, Lomb 1976; Scargle 1982; Press \& Rybicki
1989), and the phase dispersion minimization algorithm (PDM, Stellingwerf
1978).  By using three independent methods, we are less likely to be misled by
spurious results.

Using these techniques we have searched for periodic signatures from Intervals
1 through 4.  We find no significant peaks in the spectra computed from the
de-trended data from Interval 1 and 2. In Fig.~\ref{fig7} we have plotted the
results of our period search from Interval 3.  We find clear peaks detected by
all three methods at $0.647\pm0.015$ cycles/day corresponding to a period of
$1.546\pm0.038$ days.  The quoted uncertainty in the frequency is the half
width at half maximum (HWHM) of the gaussian fit to the peaks in the power
spectra. We also find weaker peaks corresponding to approximately twice the
period. In Fig.~\ref{fig8} we present light curves folded on the three biggest
peaks. The light curve folded with a period of $1.546$ days looks the
cleanest, although the light curve folded by $3.095\pm0.044$ days, which is
almost exactly double 1.546 days, looks reasonably clean as well.  The
peak-to-peak amplitude is roughly 0.15 mag although there is substantial
scatter, especially at the maximum.


The timing analysis of the data from Interval 4 also provides strong evidence
for periodic behavior (see Fig.~\ref{fig10}).  The analyses show four strong
features at $0.296\pm0.007$, $0.649\pm0.0032$, $0.652\pm0.006$, and
$1.298\pm0.006$ cycles/day, corresponding to periods of $3.370\pm0.076$,
$1.540\pm0.008$, $1.533\pm0.012$, and $0.770\pm0.0034$ days, respectively.
Here also, the strongest signal is found at $\sim0.65$ cycles/day. The $R$
band light curve from Interval 4 was folded using the four periods mentioned
above.  All of the folded light curves look reasonably clean (see
Fig.~\ref{fig11}), but the cleanest light curve is again obtained by folding
with a period of 1.540 days.  This folded light curve clearly exhibits two
maxima and minima per phase, and the peak-to-peak amplitude is $\sim 0.36$
mags, more than double the value corresponding to Interval 3.


The periodic behavior could be due to several different underlying
mechanisms. During outbursts, modulations in the light curve can arise from
the precession of the tidally distorted accretion disk, known as superhumps
(Whitehurst 1988), or from the X-ray heated secondary star (van Paradijs \&
McClintock, 1995).  During quiescence, if the light curve has periodic
features, it is usually an ellipsoidal variation.  Ellipsoidal light curves
exhibit two equal maxima and two unequal minima when folded by the orbital
period (van Paradijs \& McClintock 1995). During the transition from outburst
to quiescence optical modulations can be caused by a combination of all of
these effects.

The presence of disk heating can explain the absence of a second minimum in
the folded light curve from Interval 3 (see Fig.~\ref{fig8}).  As the source
was still in outburst during this time, secondary heating from the disk may
have been significant enough to produce a broad maximum during superior
conjunction, which results in a light curve with only one minimum and one
maximum per orbit (Orosz 1996).  This scenario is supported by the fact that
the peak to peak amplitudes of the optical light curve increased, by a factor
of $\sim 2$, as the overall brightness continued to drop between Interval 3
and 4.  This can only happen if the modulation was coming from the secondary
star and not from the disk itself.  Therefore, we favor the interpretation of
the Interval 3 and 4 light curves as a combination of secondary heating and
ellipsoidal modulation, which together suggest an orbital period $\sim 1.54$
days.

An alternative interpretation of Interval 3 data using superhumps seems to be
ruled out by our observations.  Superhump periods are usually $\sim 1\%$
longer than the orbital period, hence the combined, detrended Interval 3 and 4
light curves should manifest a phase offset when folded by the the same
period, if the Interval 3 periodicity was from a superhump.  To test this
possibility, we combined the de-trended data from Interval 3 and 4 and
performed timing analysis of this joint data set and found a strong peak at
$P=1.541\pm0.009$ days, or frequency of $0.649\pm0.004$ cycles/day
(Fig.~\ref{fig12}).  The combined data set was folded with 1.541 days and is
shown in Fig.~\ref{fig13}.  As the maxima and minima line up in phase,
superhumps appear to be ruled out because a $1\%$ difference in period would
have caused an appreciable offset in phase over the span of the observation.
The existing evidence shows that the true orbital period is close to 1.54 days
and that ellipsoidal variability is present.


\section{Correlation with the X-ray Light Curve}
During the entire outburst, RXTE PCA and ASM observations were conducted on a
daily basis. The detailed analysis of the timing and spectral analysis are
presented elsewhere (see Paper I,II,IV).  Here we compare the optical light
curve from 1998, as well as from Interval 1 and 2 with the X-ray light
curve. In particular, we will make use of the detailed spectral information
from the PCA data to decompose the total X-ray flux from 2-20 keV into the
contribution from the model components.  The model used in Papers I and IV
consists of the following components: interstellar absorption, multi-color
disk blackbody, power law, and a broad Fe emission line.  Therefore, based on
the fit using the model above, the total X-ray flux can be decomposed into the
disk blackbody flux and the power law flux.

From Figs.~\ref{fig14} and \ref{fig15} it is clear that the shape of the
optical and X-ray light curves are quite different.  For example, during the
steep decline of the 1998 optical light curves between MJD 51080 and 51100,
the X-ray disk flux was nearly constant, until MJD 51100 (see
Fig.~\ref{fig14}) when it suddenly increased.  During this time the power law
flux decreased very slowly with an e-folding time of $\sim 63.0$ days, nearly
double the optical value.  The slow decline in the X-ray intensity is unusual,
since typically the X-ray decay is much faster than the optical -- compare the
average e-folding decay time-scales of 17.4 and 67.6 days for the X-ray and
optical light curves, respectively, computed from a large sample of SXTs (Chen
et al.\ 1997).


During Interval 1, the earliest stages of the re-flare, the optical flux
slowly declined as the X-ray flux first increased slightly and then decreased
by a similar amount, thereby producing a concave X-ray light curve with a
maximum at around MJD 51200.  During Interval 1 there were two prominent dips
in the $V$ light curve at MJD 51200.3 and 51215, with amplitudes of $\sim$0.32
\& 0.25 respectively (see the Observations section). The disk flux light curve
shows no sign of similar dips (see Fig.~\ref{fig15}). On the other hand, the
power-law flux is anti-correlated with the first optical dip, exhibiting a
mini-flare, although there is no response to the second optical dip.  The
power-law mini-flare peaks at MJD 51201.78, which is nearly 2 days after the
peak of the optical dip, and has an X-ray flux of roughly $1.0\times10^{-8}$
erg s$^{-1}$ cm$^{-2}$ relative to the ten day average between MJD 51190 and
51200. The relative amplitude of the power-law mini-flare is roughly twice as
large as the corresponding optical dip.  Since we only have two data points
representing this mini-flare, the estimate of the peak could be off by as much
as half a day.

At the beginning of Interval 2, MJD 51230, the optical light curve decreases
sharply, by $\sim 0.6$ magnitudes.  Again, the disk flux showed no change as
the optical flux dipped.  However, the power-law flux light curve is once
again anti-correlated with the $V$-light curve. It is difficult to determine
when exactly the power-law flux began to increase, due to the large error bars
and sparse data coverage.  Clearly, by MJD 51232.25, the power-law flux had
increased by $4.42 \times 10^{-9}$ erg s$^{-1}$ cm$^{-2}$, compared to the
average values between MJD 51226 to 51230.  Since the optical flux during
outburst is typically believed to be dominated by the component from the
accretion disk, the absence of any correlation between the optical flux and
the X-ray disk flux is surprising.  On the other hand, our results suggest
that both the optical and the power-law component may be a manifestation of
the same physical process.


One possibility is that the optical flux is upscattered to the X-ray by the
hot corona. In this scenario, during the optical dip, a larger fraction of the
optical photons are reprocessed into hard X-rays by compton scattering in the
corona, resulting in a decrease in the optical photons accompanied by an
increase in the power-law component.  This can occur if there is an increase
in the size of the corona or an increase in the solid angle it subtends at the
outer edge of the disk, or if the opacity suddenly increases.  If variations
in the properties of the outer edge of the accretion disk were causing the
optical dips (e.g.  due to a mass transfer instability), we should expect to
see a corresponding dip in the X-ray as well, with a delay set by the viscous
timescale. The viscous timescale is much longer than the $\sim 2.25$ days
delay we find, hence this possibility seems to be ruled out by the data.
Nevertheless, even if the optical flux is up-scattered to X-ray, the delay of
$\sim 2$ days is difficult to explain since the corresponding light travel
time far exceeds the size of the compton cloud. Perhaps there is a time
dependence in the efficiency of the scattering resulting in the delay, but it
is unclear how this may arise.
 
Finally, we see remarkable transitions in both the disk flux and power-law
light curves starting at around MJD 51240 (see Fig~\ref{fig15}). During the
next ten days, the disk component dropped by $3.5\times10^{-8}$ erg s$^{-1}$
cm$^{-2}$, or by a factor of 3.0 and the power-law flux increased by nearly
$3.80\times10^{-8}$ erg s$^{-1}$ cm$^{-2}$, or a factor of 4.9.  As mentioned
before, between MJD 51230 and MJD 51245 there are large fluctuations in the
optical light curve, complicating the comparison of the X-ray and optical
light curves.  Clearly the e-folding timescale of the optical light curves
between Interval 1 and 2 are different, but it is not clear whether the
transition occurred at around MJD 51230 or MJD 51240 (see Fig.~\ref{fig15}).
It is clear, however, that the disk flux between MJD 51240 and 51250 is now
correlated with the optical whereas the power-law component is
anti-correlated.  The correlation of the disk flux and anti-correlation of the
power-law component with optical magnitudes between MJD 51240 and 51250
suggests the onset of a global change in the accretion disk spanning the outer
and inner disk as well as the corona.

\section{Discussion}

Recent progress has been made in understanding the optical light curve and
their relation to the X-ray light curve for a number of well known black hole
SXTs.  By using a model consisting of an inner ADAF region and an irradiated
outer thin disk, Esin et al. have fitted the optical and X-ray light curves of
A0620-00, a canonical black hole SXT, and GRS 1124-68, also known as Nova
Muscae (see Esin et al.\ 1997; hereafter EMN, and EKMN).  The model
essentially depends only on two key parameters, the accretion rate in
Eddington units, $\dot{m}$, and the transition radius between the ADAF and the
thin disk, $r_{tr}$.  The optical flux, which originates from the outer
region, is dominated by X-ray irradiation of the disk surface, and therefore
is proportional to the X-ray flux emitted from the inner regions (van Paradijs
1996; King \& Ritter 1998).  It is clear that the particular choice of
$\dot{m}$ and $r_{tr}$ will dictate not only the X-ray spectrum but the
optical spectrum as well, due to the coupling by X-ray heating.  According to
the scenario outlined in EMN and EKMN, after the source has reached its
maximum brightness, the accretion rate $\dot{m}$ begins to decrease
exponentially in time as the transition radius $r_{tr}$ increases, although
not necessarily at the same time.  At the highest accretion rate and smallest
transition radius, the source is in the {\it very high state}.  Subsequently,
as $\dot{m}$ decreases, the source passes through the {\it high state} and
at a  critical value of at $\dot{m}$, reaches the {\it intermediate state}
when 
the inner disk starts to evaporate.
During the {\it intermediate state}, the spectral and temporal
characteristics shift from the {\it high state} to the {\it low
state}. Finally, the outer accretion disk is restricted to large radii and the
source enters the {\it quiescent state}. 
Spectral transitions, as described above, occurred in XTE J1550-564 throughout
the outburst and re-flare (Paper IV; Fig~\ref{fig14}).  From the very
beginning of the outburst to MJD 51115, the object was in the {\it very high
state}.  Between MJD 51115 and 51150 the source faded rapidly and entered the
{\it high state}, although occasionally the source appeared to be in the {\it
intermediate state}. During most of the the re-flare (after MJD 51150) the
source was in the {\it high state} although the source briefly entered the
{\it very high state} after MJD 51230, at which time the power-law flux
increased considerably.  Finally the spectrum of one of the last few
observations resembles the {\it low state} (see Fig.~\ref{fig14}).  The
luminosity of the source was consistent with the scenario laid out by
EMN, where the {\it very high state} is more luminous than the {\it high
state} which is in turn more luminous than the {\it intermediate state} and
{\it low state}, with the exception of the last few observations which is
perhaps too bright to be in the {\it low state}.

Although the scenario laid out by EMN and EKMN applies very well to A0620-00
and Nova Muscae, it is rather difficult to understand the light curves of XTE
J1550-564 within this framework.  The obvious difficulty lies in the
dramatically different morphologies of the optical and X-ray light curves.  We
also find substantial optical emission during Interval 3, even though there is
no discernible X-ray emission.  Clearly such behavior can not be explained
within the context of a simple irradiated disk, as the optical flux is
proportional to the X-ray emission in such models (van Paradijs \& McClintock
1995; King \& Ritter 1998).  Moreover, there are substantial features in the
optical light curve, such as sudden changes in the e-folding timescale, dips
and mini-flares, which are absent in the X-ray light curve.

Recently ELH have modeled the optical, UV, and X-ray light curves of GRO
J1655-40 during the 1996 outburst, which are quite similar to the light curves
of XTE J1550-564 from Intervals 1 and 2. Both sources exhibit an exponentially
declining optical light curve, whereas the X-ray remains constant or increases
slightly.  Furthermore, both outbursts began as a flat top with a {\it very
high state} spectra, followed by a re-flare.  The model depends on several
critical assumptions.  First, the mass transfer rate $\dot{M}_{T}$ must be
close to the stability limit for an irradiated disk.  Second, $\dot{M}_{T}$
needs to be increased above the critical accretion rate by the irradiation of
the secondary due to the X-ray emission from the disk. This prevents the X-ray
flux from decreasing exponentially. Finally, the warping of the disk required
by recent theoretical work for the outer disk to be irradiated (Dubus et al.\
1999) must diminish with a characteristic warp damping time-scale
$t_{damp}\sim4\alpha^2t_{vis}$ (ELH) which is about 100 days for $\alpha=0.1$,
where $t_{vis}$ is the the viscous time-scale and $\alpha$ is the parameter of
viscosity (Shakura \& Sunyaev 1973).  Due to the incomplete optical coverage
of GRO J1655-40, the re-flare which occurred $\sim 270$ days after the initial
outburst was not considered at all in ELM.  Although this scenario may apply
to the optical light curves of XTE J1550-564, judging from the similarity of
the source and the light curves, the present data set clearly requires a
detailed consideration of the re-flare.

There has been great attention paid to the exponential nature of optical and
X-ray light curves, perhaps at the expense of recognizing the interesting
exceptions.  This is not surprising, as optical data coverage is often
incomplete and indeed the prominent X-ray light curve morphology for SXTs is
the fast rise followed by an exponential decay (or FREDs) (Chen et al.\ 1997).
Disk instability models generally produce such exponentially decaying light
curves (Mineshige \& Wheeler 1989; Cannizzo, Chen, \& Livio 1995; Hameury et
al.\ 1998).  The present data of XTE J1550-564 provide a nearly complete
picture of the entire optical outburst including the re-flare, and thus pose a
qualitatively more complex challenge to models of outburst and decay of SXTs.

\section{Summary}
We have presented comprehensive optical light curves for XTE J1550-564 from
September 1998 to August 1999 and compared them with the X-ray results
presented in Papers I, II, and IV.  The shape of the optical and X-ray light
curves differ substantially from the typical fast rise and exponential decay
and the correspondence between the overall optical and X-ray flux levels is
poor, although there are several instances of anti-correlation between the
optical flux and the hard power law component of the X-ray system.  Analysis
of the R-band data at and near quiescence reveal strong periodic signatures at
$\sim 1.54$ days with peak to peak amplitudes of 0.15-0.36 mags which we
interpret as the orbital period.  The single hump behavior in the decay toward
quiescence may be due to ellipsoidal modulation and X-ray heating, whereas the
double humped signal in quiescence is presumably due to ellipsoidal
modulation, with marginal irradiation from the disk.  However, more data from
quiescence will be required to fully resolve the nature of the modulation.

\acknowledgements We thank the two YALO observers, David Gonzalez Huerta and
Juan Espinoza, for providing data in a timely manner. We would like to thank
S. Tourtellotte and E. Terry for their assistance with data reduction and
J. Greene for her assistance with data acquisition.  We thank our referee, Ann
Esin, for her useful comments and suggestions. R.J. would like to thank
M. Garcia, K. Menou, M. Nowak, and M. Sako for useful discussions.  Financial
support for this work was provided by the National Science Foundation through
grant AST 97-30774.

\clearpage
\figcaption[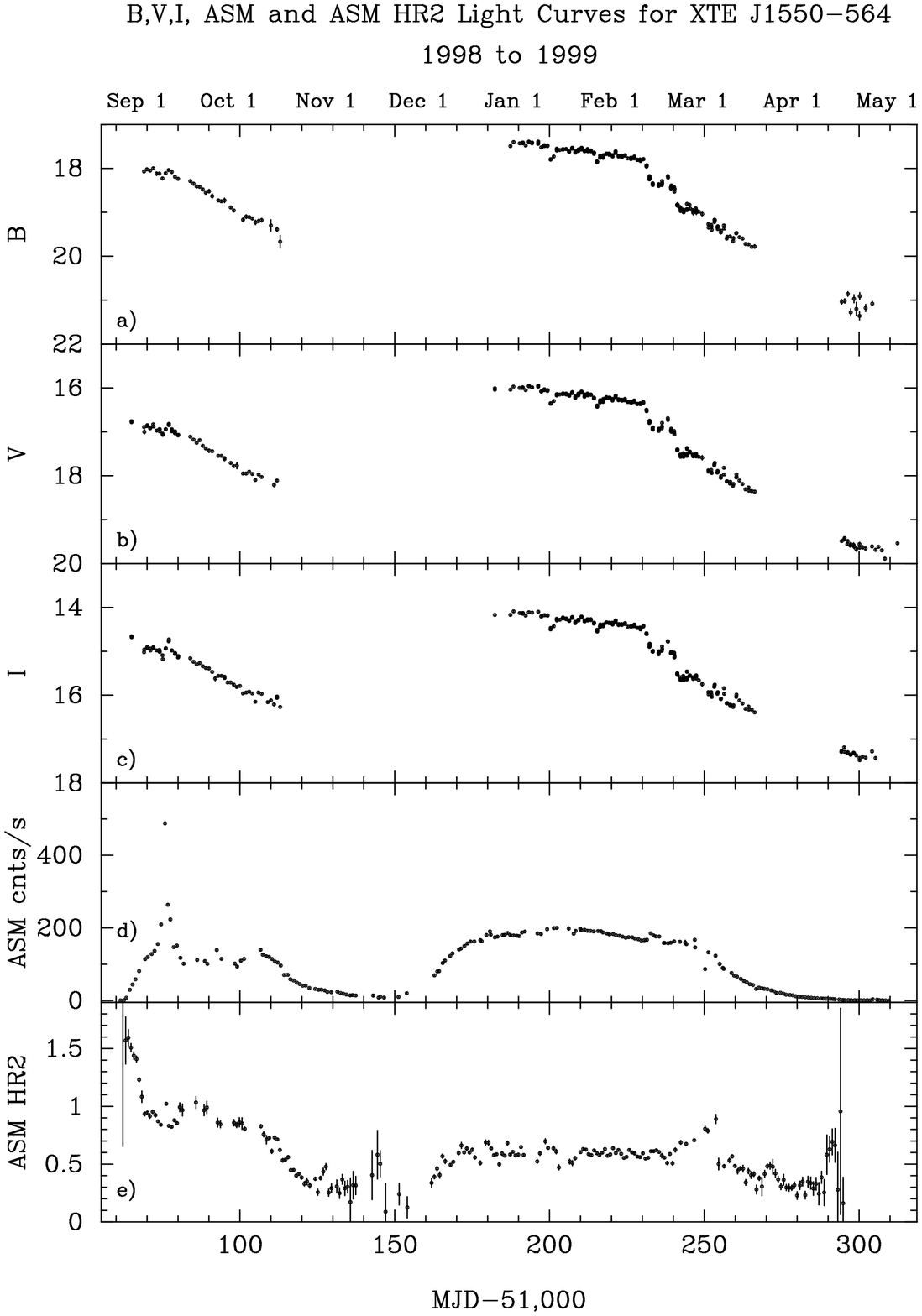]{ Top to Bottom (a,b,c,d,e): $B$,
$V$, $I$, ASM and ASM HR2 light curves covering the entire observation
campaign (including what was presented in Paper III).  ASM HR2 (hardness ratio
2) is defined as the ratio of ASM count rates between 5-12 and 3-5 keV.  HR2
values were not computed beyond day MJD 51,295 since the ASM count rate is too
low to obtain meaningful ratios.  The calendar dates are indicated on the top
of the plot.  The corresponding modified Julian date (MJD) is denoted by the
first character of the months, i.e. the labels are left justified with respect
to the corresponding MJD.
\label{fig1}
}

\figcaption[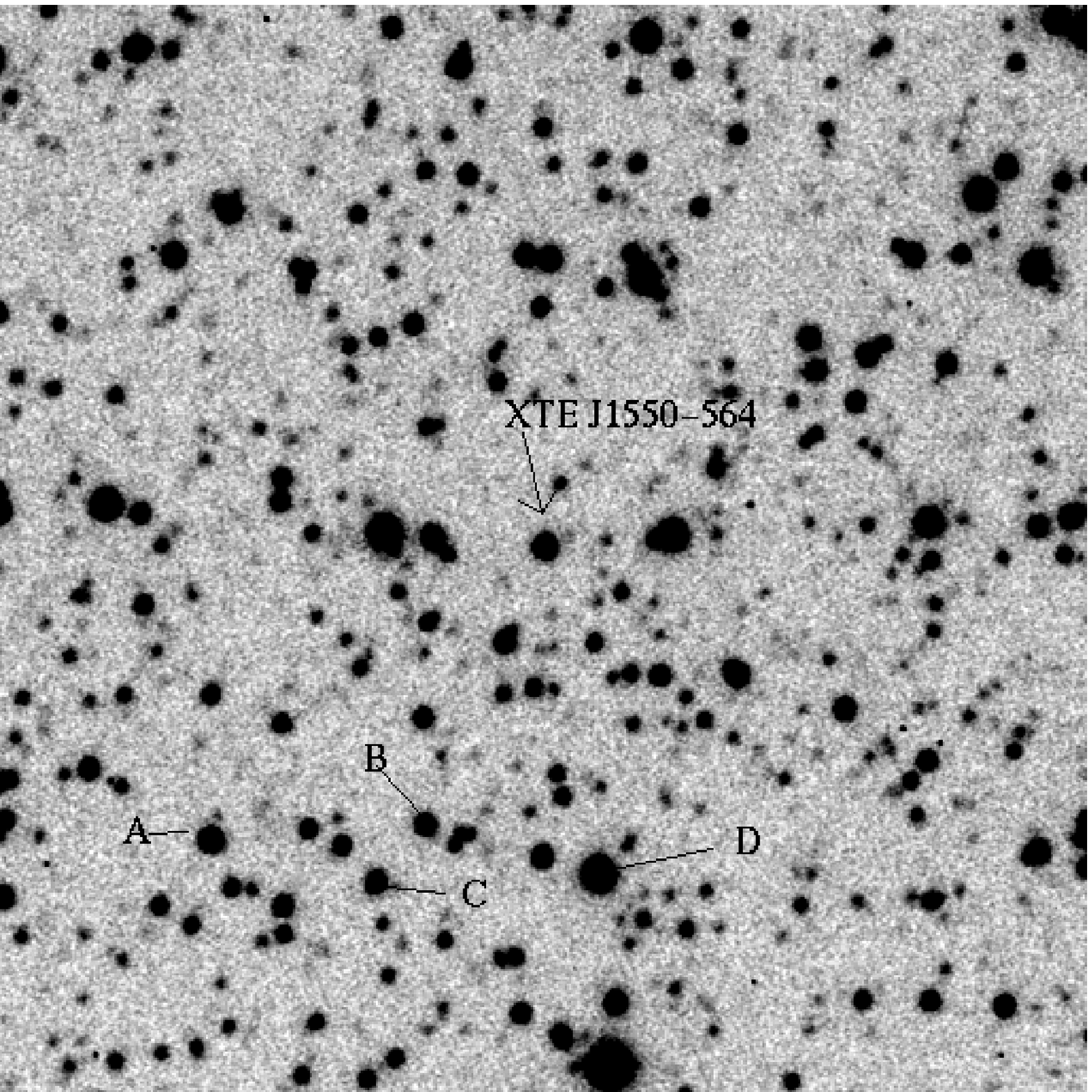]{ Finding chart of XTE J1550-564 taken with the YALO 1m
telescope (V $3 \arcmin \times 3 \arcmin$).  North is to the right and east is
to the bottom.
\label{fig0}
}

\figcaption[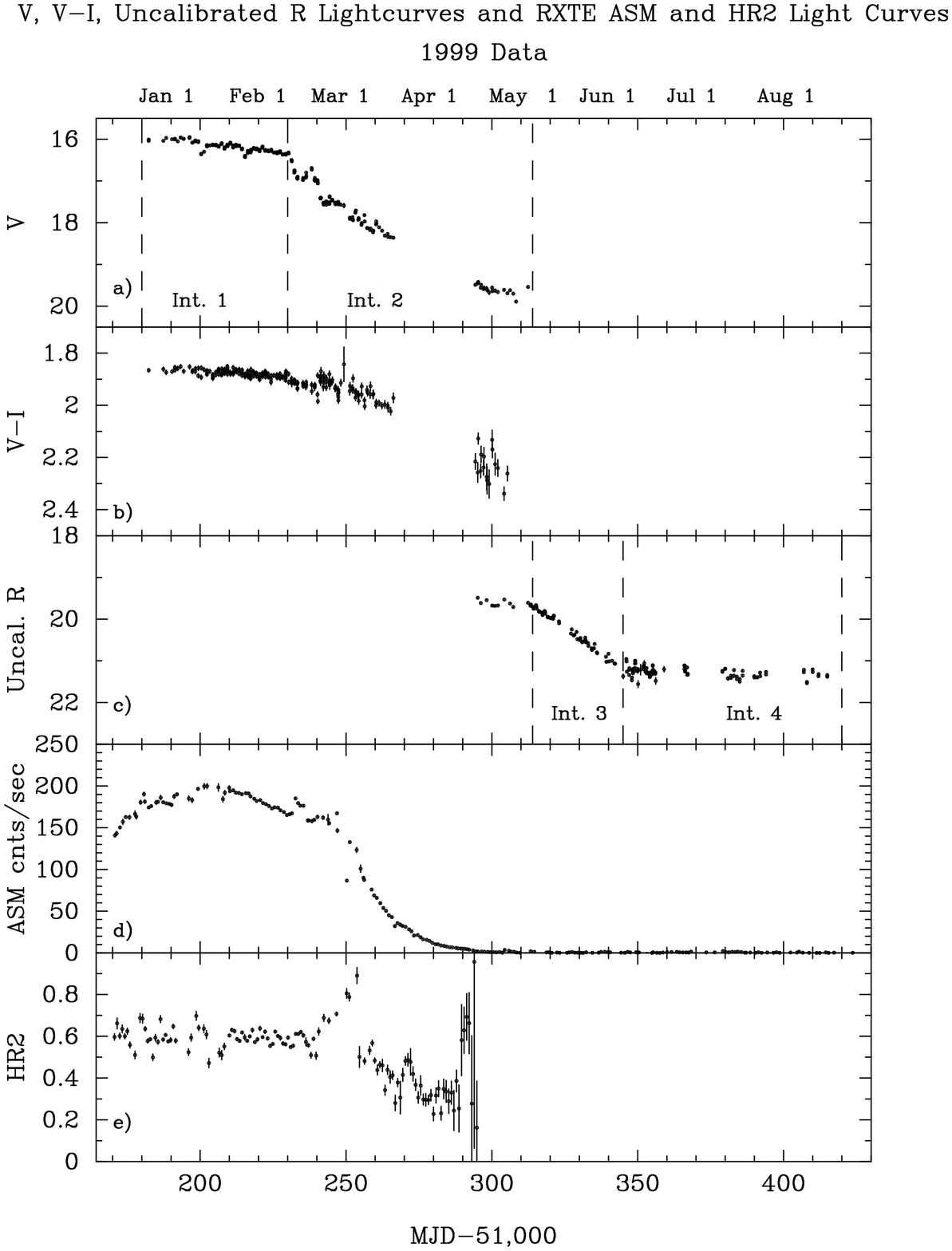]{ Top to Bottom (a,b,c,d,e): $V$,
$V-I$ color, $R$, ASM and ASM HR2 light curves for the 1999 re-flare.  The
calendar dates are indicated on the top of the plot.  The corresponding MJD is
denoted by the first character of the months, i.e. the labels are left
justified with respect to the corresponding MJD. The light curve is divided
into four intervals as indicated.
\label{fig2}
}

\figcaption[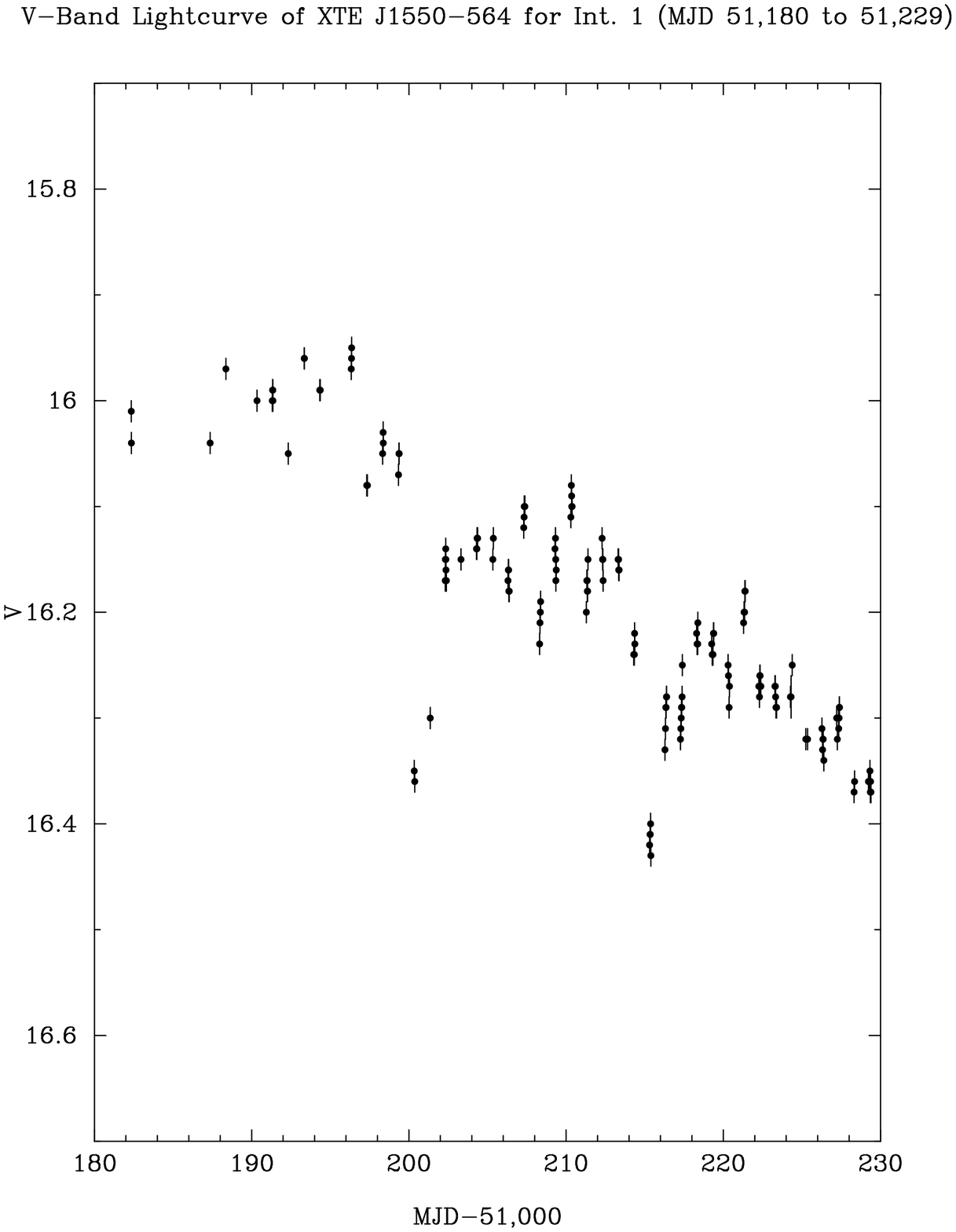]{ The $V$ light curve from Interval 1 (MJD 51180 to
51229). Note the prominent dips and the slow decline, with an e-folding time
of 124 days.
\label{fig3}
}

\figcaption[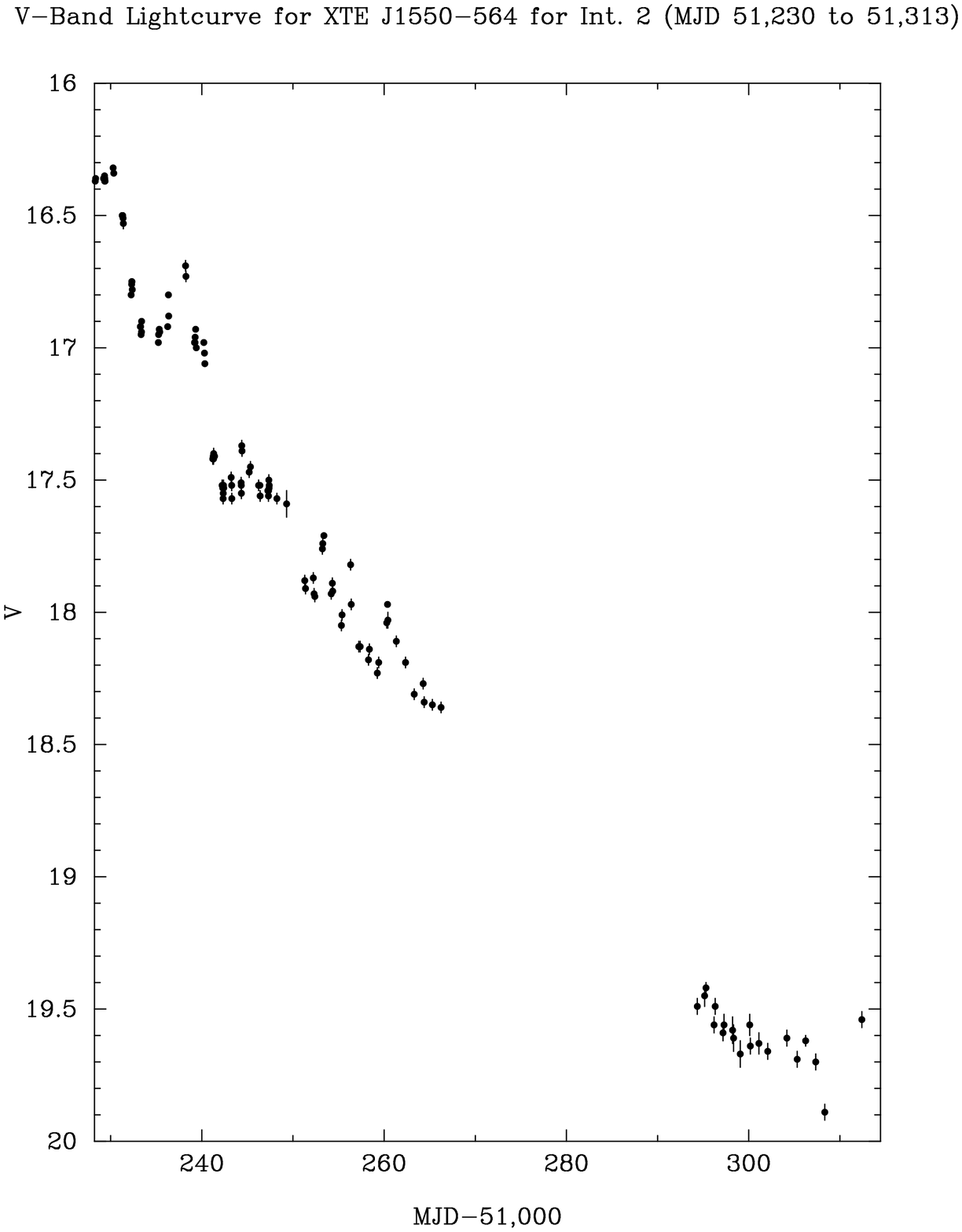]{ The V light curve from Interval 2 (MJD
51230 to 51312). The e-folding time for the data from
MJD 51244 to 51300 is 27 days.
\label{fig4}
}

\figcaption[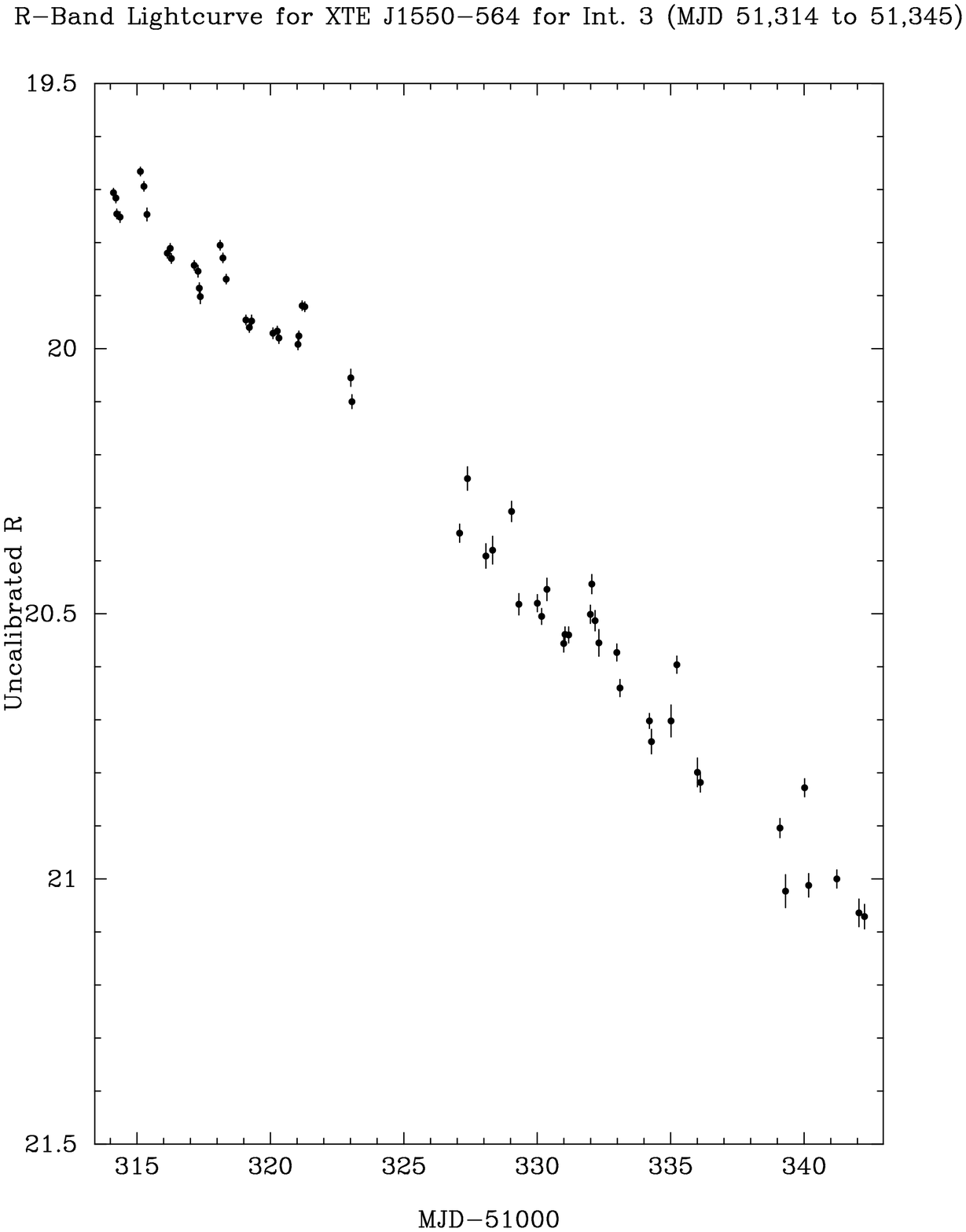]{ The $R$ light curve from Interval 3 (MJD 51313 to
51345).  We find evidence for periodic behavior from this interval.  The
e-folding time is 22.6 days.
\label{fig5}
}

\figcaption[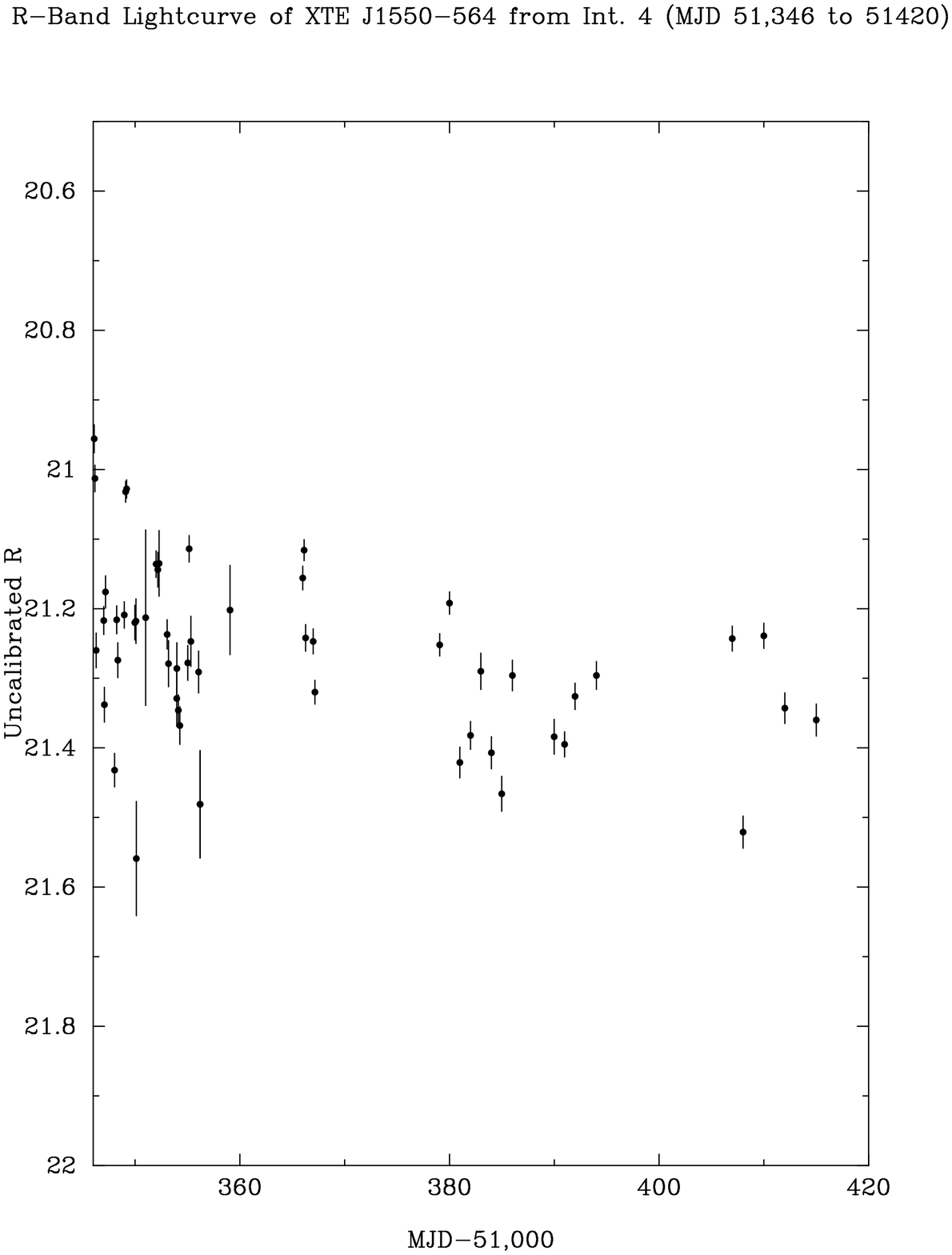]{ The $R$ light curve from Interval 4 (MJD 51346 to
51420). We find periodic motion from this interval as well.  There is a slowly
decaying component to the light curve corresponding to a 
very long e-folding time of 362 days.
\label{fig6}
}

\figcaption[fig8.ps]{ Result of the three different
period searching algorithms for Interval 3 data. From top to bottom: PDM
(Phase Dispersion Minimization) statistics, CLEAN power spectrum, and
Lomb-Scargle power spectrum.  The strongest signals is found at
$0.647\pm0.015$ cycles/day or a period of $1.546\pm0.038$ days.  We also find
peaks in the PDM power spectrum at $0.323\pm0.005$ cycles/day and in the
Lomb-Scargle periodogram at $0.356\pm0.015$ cycles/day, corresponding to
periods of $3.095\pm0.044$ and $2.809\pm0.03$ days.
\label{fig7}
}

\figcaption[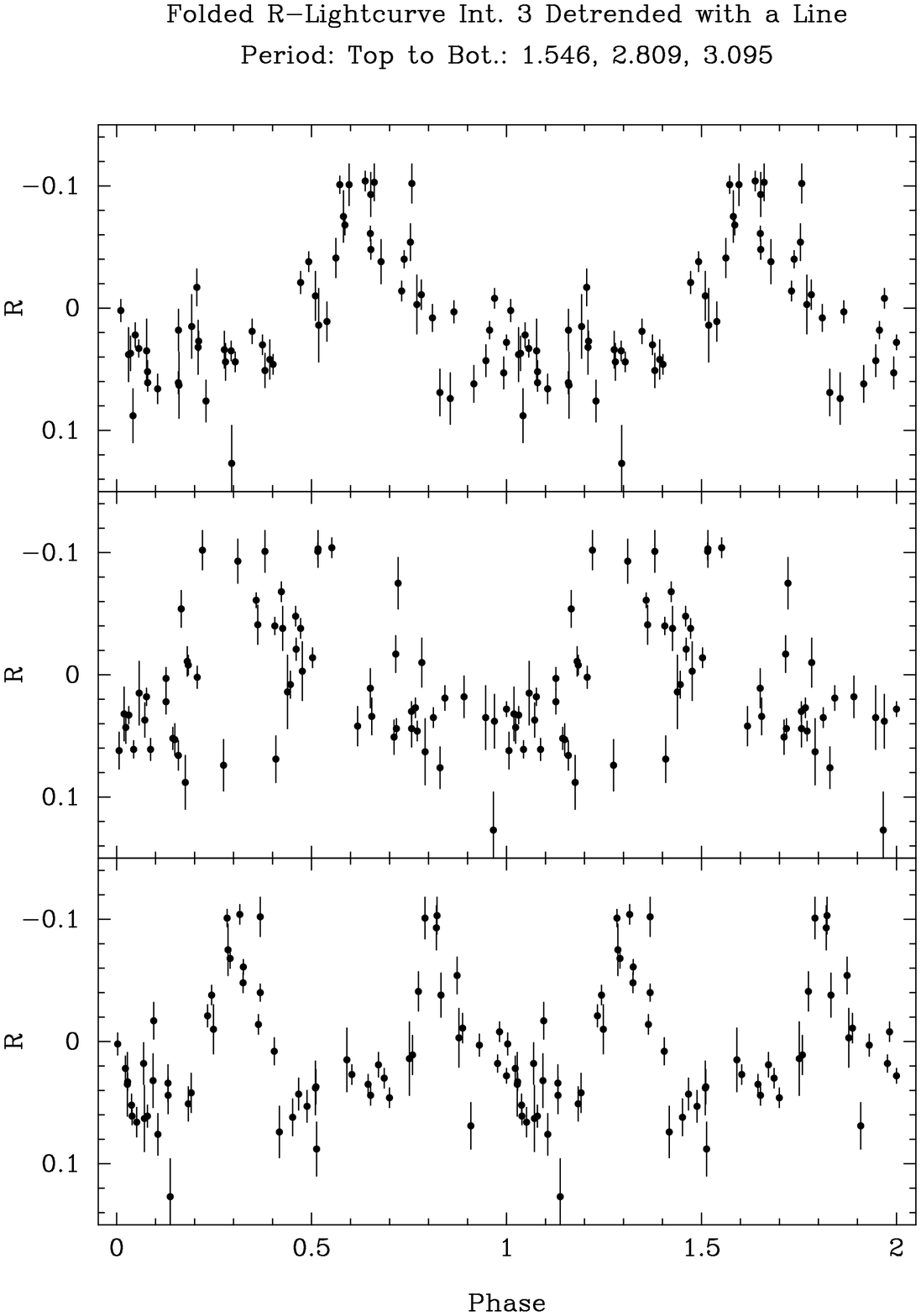]{ De-trended folded R-light
curves from Int. 3. Top: The folding period is 1.546 days. The folded light
curve is clearly periodic and sinusoidal.  Middle: The light curve folded with
a period of 2.809 days.  Bottom: The light curve folded with a period of 3.095
days. This is nearly double the first folding period, hence we see two cycles
per phase.
\label{fig8}
}

\figcaption[fig10.qdp]{ Result of the three
different period searching algorithms for Interval 4.  From top to bottom PDM:
(Phase Dispersion Minimization) statistics, CLEAN power spectrum, and
Lomb-Scargle power spectrum.  The strong signals are found at $0.296\pm0.007$,
$0.649\pm0.0032$, $0.652\pm0.006$, and $1.298\pm0.006$ cycles/day
(corresponding to periods of $3.370\pm0.076$, $1.540\pm0.008$, $1.533\pm0.012$
and $0.770\pm0.0034$ days).
\label{fig10}
}

\figcaption[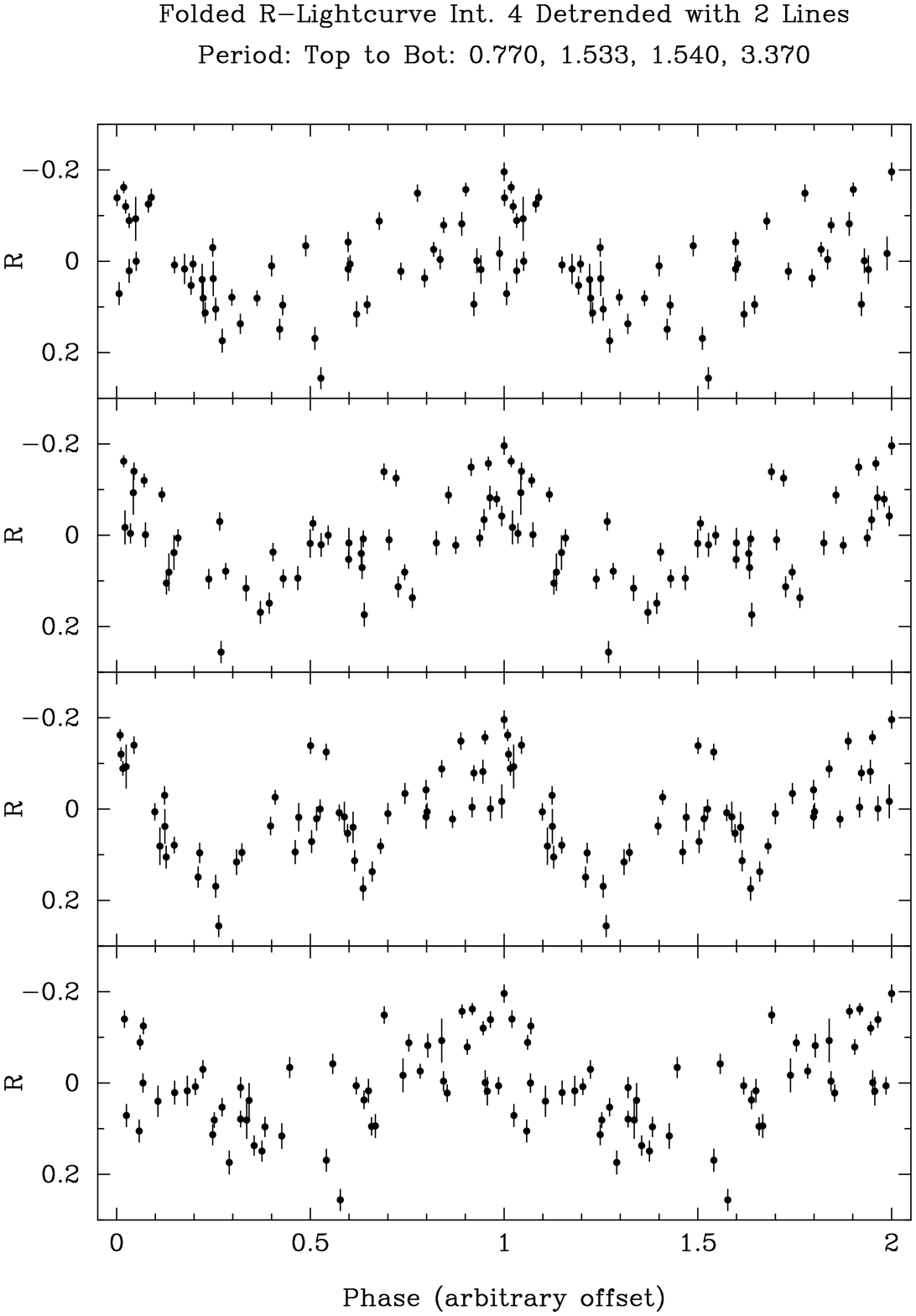]{ De-trended folded light curves
from Int. 4, folded with periods corresponding to periods of 0.770, 1.533,
1.540, and 3.370 (from top to bottom).  The light curve folded at $P=1.540$
days exhibits a double humped sinusoidal behavior, although there is
considerable noise.
\label{fig11}
}

\figcaption[fig12.ps]{ Result of three different
period searching algorithms for the combined the de-trended data from Int. 3
and Int. 4. From top to bottom: PDM (Phase Dispersion Minimization)
statistics, CLEAN power spectrum, and Lomb-Scargle power spectrum. The
strongest feature is found at $f=0.649$ cycles/day or $P=1.541$ days.
\label{fig12}
}

\figcaption[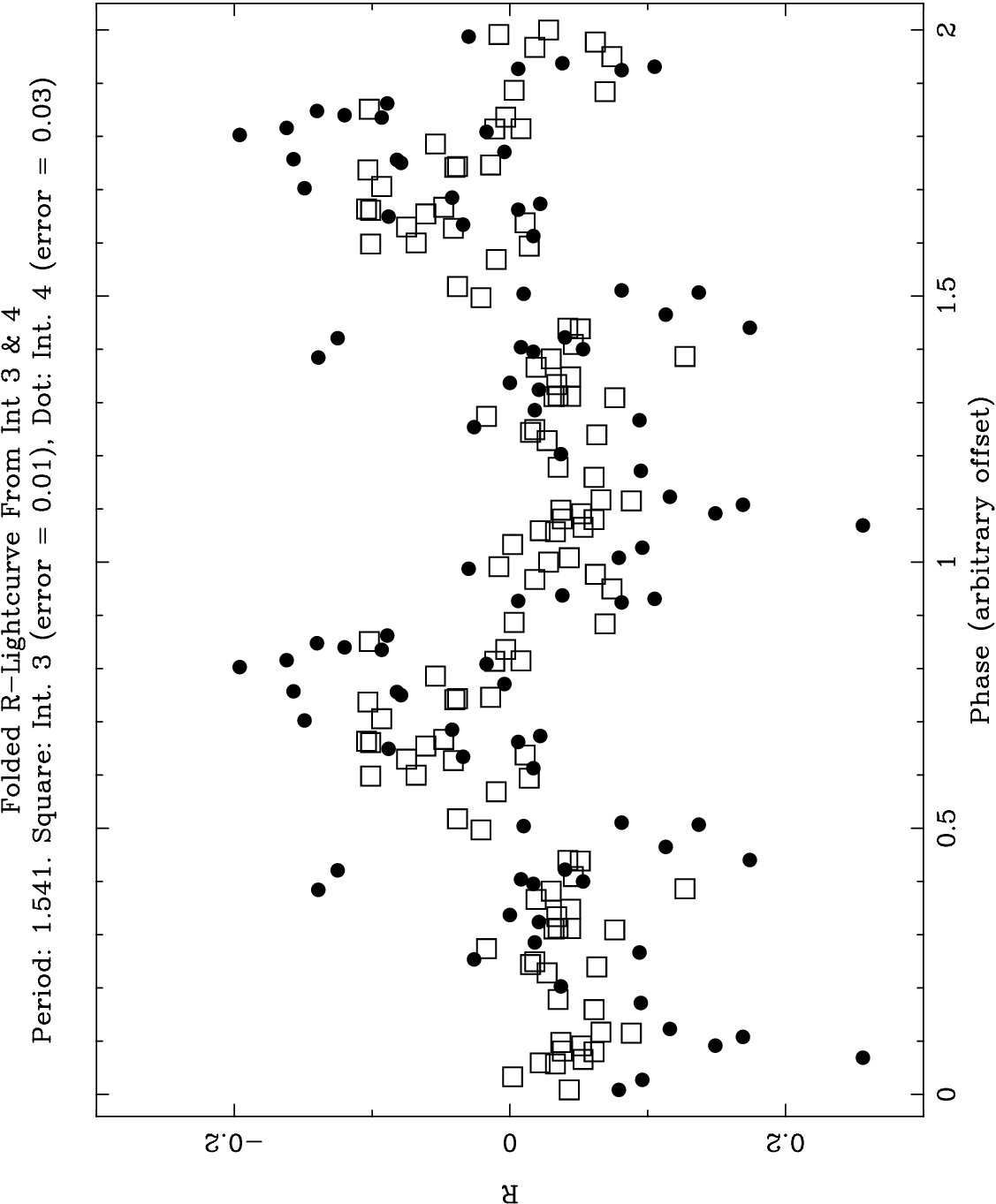]{ De-trended folded $R$ light
curve from Interval 3 and Interval 4 ($P= 1.541$ days). Data from Intervals 3
and 4 were separately de-trended.  Squares correspond to data from Interval 3
and the dots correspond to Interval 4. The phase seems to be maintained
throughout both intervals.
\label{fig13}
}

\figcaption[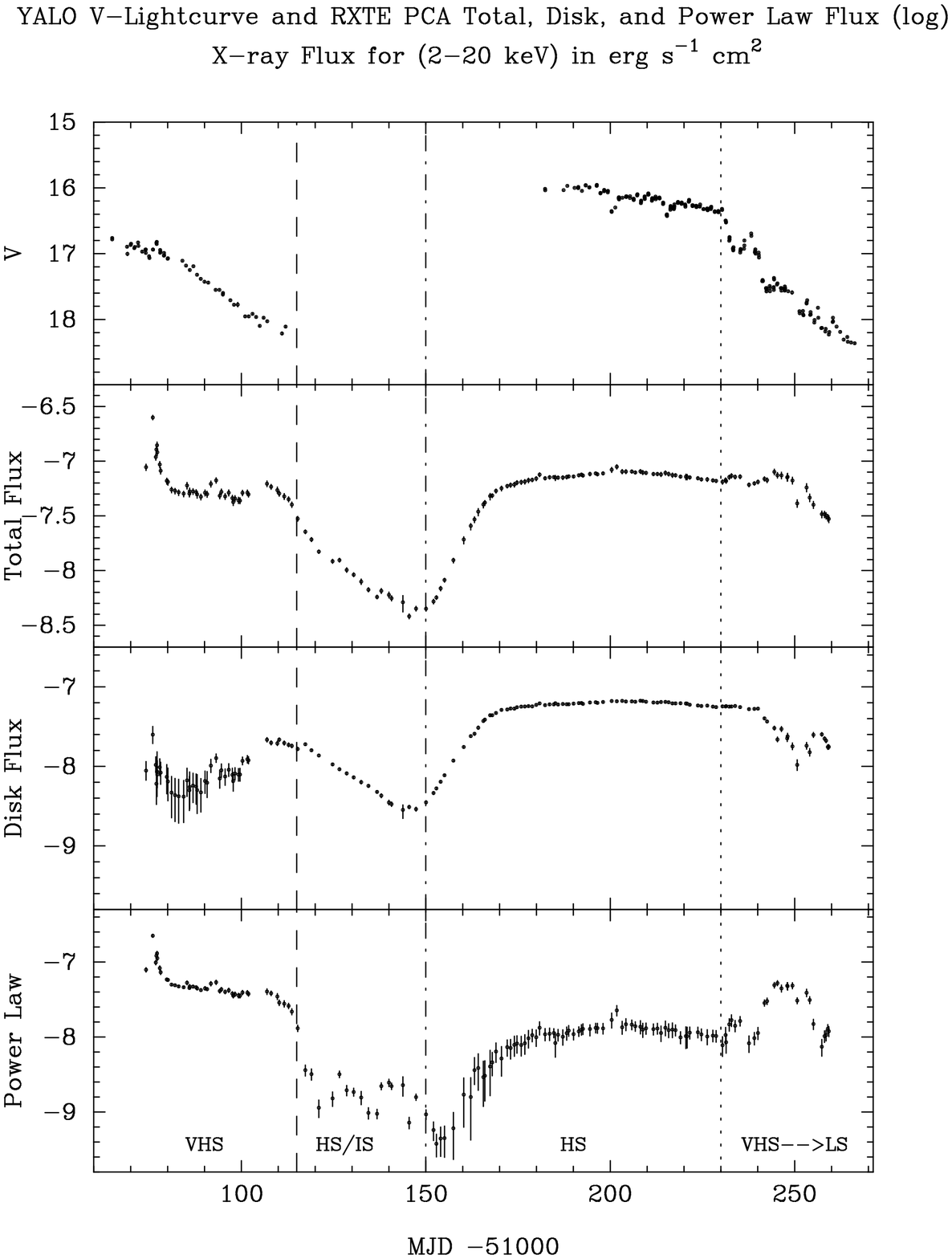]{ $V$ light curve and the logarithm
of the X-ray fluxes.  The X-ray fluxes (2-20keV) are computed from the best
fitting model (see Paper I \& IV).  From top to bottom: The $V$ band light
curve, the total X-ray flux, disk flux, and the power law flux light curves.
The vertical lines enclose regions with different spectral states, which are
indicated in the lowest panel.  The source was in the {\it very high state}
(VHS) from the beginning of the outburst until MJD 51115. Subsequently between
MJD 51115 and 51150, the source entered the {\it high state} (HS), although
occasionally the source appeared to be in the {\it intermediate state}
(IS). During most of the re-flare (after MJD 51150) the source was in the {\it
high state}.  At the very final stage of the re-flare (beyond MJD 51230) the
power-law flux increased and the source temporary entered the {\it very high
state} before entering the {\it low state} (LS).
\label{fig14}
}

\figcaption[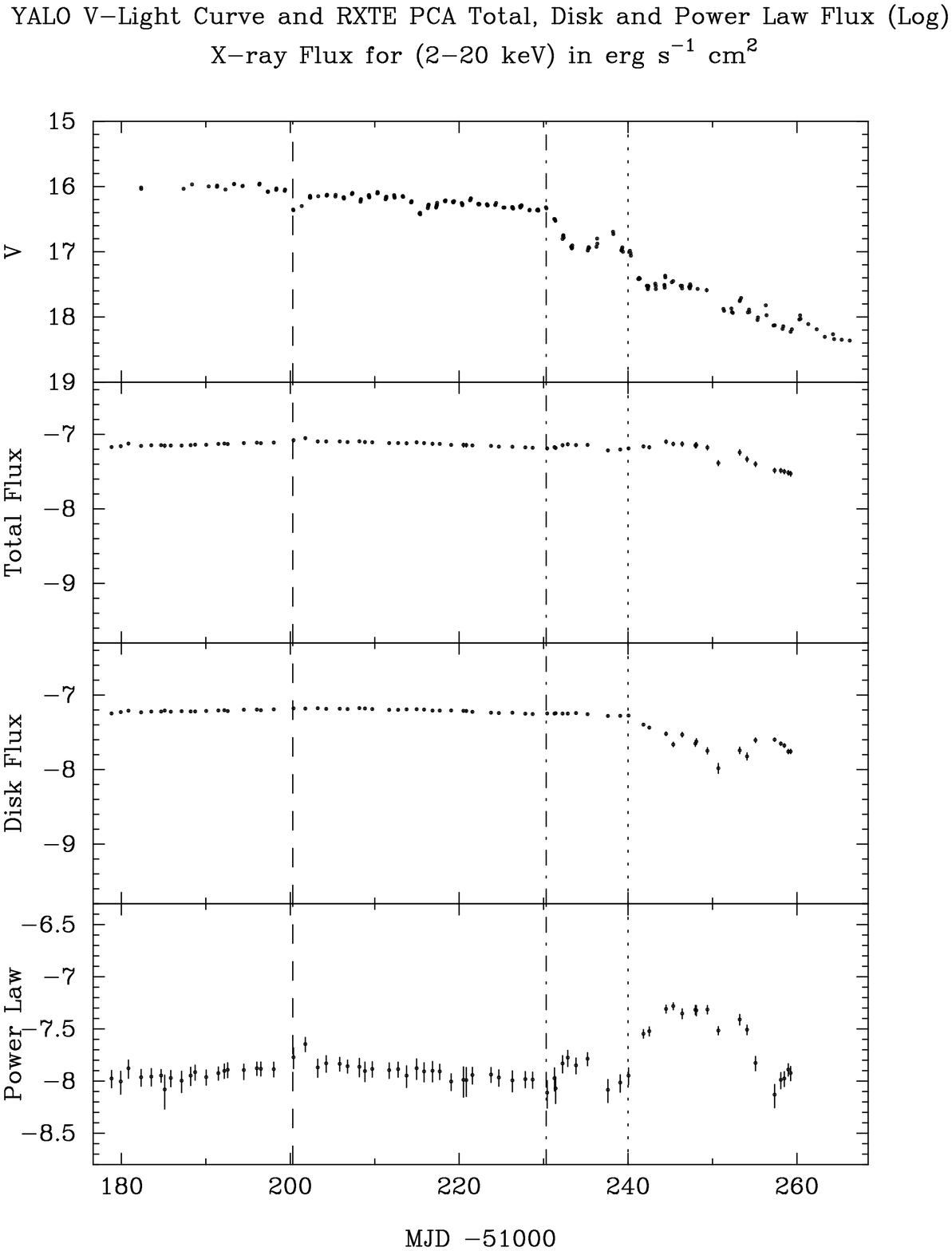]{ The same light curves as
Fig.~\ref{fig14} zoomed on Intervals 1 \& 2.  The dashed line corresponds to
the optical dip at time 200.3, the dash-dotted line corresponds to the end of
Interval 1 and the beginning of Interval 2 and finally the dotted line
corresponds to the onset of dramatic global changes in the accretion disk,
exemplified by the concurrent changes in the optical flux as well as the disk
and power law fluxes.
\label{fig15}
}


\begin{deluxetable}{cccc} \footnotesize  \tablecaption{Journal of
 Observations \label{tab:journal}}
\tablehead{ \colhead{} & \colhead{Date MJD, (UT MM/DD)\tablenotemark{a}} & \colhead{Number of
Exposures} & \colhead{Exposure Times (s)} } 

\startdata
$B$ & $51187.37 \sim 51266.27,\ (01/09 \sim 03/29)$ & 230 & 300, 600, 900\tablenotemark{b} \\ 
  & $51294.37 \sim 51304.26,\ (04/26 \sim 05/06)$  & 11  & 1200s \\ 
\cline{2-4} \nl
$V$ & $51182.34 \sim 51266.25,\ (01/04 \sim 03/29)$  & 260 & 300, 450\tablenotemark{c} \\ 
  & $51294.36 \sim 51312.40,\ (04/26 \sim 05/14)$  & 20  & 900, 1200
  \tablenotemark{d} \\
\cline{2-4} \nl
$I$ & $51182.35 \sim 51266.26,\ (01/04 \sim 03/29)$  & 245 & 300,
450\tablenotemark{e} \\
  & $51294.24 \sim 51305.33,\ (04/26 \sim 05/07)$  & 17 & 900 \\
\cline{2-4} \nl
$R$ & $51295.18 \sim 51367.16,\ (04/27 \sim 07/08)$\tablenotemark{f} &136 & 900,
1200\tablenotemark{g} \\
  & $51379.05 \sim 51415.00,\ (07/20 \sim 08/24)$ & 34 & 1200 \\
\enddata

\tablenotetext{a} {All observations were taken in 1999}
\tablenotetext{b}{ Switched to 600s after 01/26 and to 900s after 03/23}
\tablenotetext{c,e}{Used 450s from 03/23 to 03/29}
\tablenotetext{d} {Three exposures between 05/08 and 05/14 in 1200s}
\tablenotetext{f} {There is a gap between 06/30 and 07/06}
\tablenotetext{g} {Switched to 1200s after 05/20}
\end{deluxetable}

\clearpage

\begin{deluxetable}{cccc} \footnotesize  \tablecaption{Calibrated Standards 
\label{tab:mag}} 
\tablehead{ \colhead{Star ID} & \colhead{$B$} & \colhead{$V$} & 
\colhead{$I$} } 
\startdata
A & $18.01\pm0.03$ & $16.13\pm0.02$ & $14.13\pm0.03$ \\
B & $18.77\pm0.06$ & $16.96\pm0.02$ & $14.89\pm0.03$ \\
C & $19.11\pm0.08$ & $16.84\pm0.02$ & $14.21\pm0.03$ \\
D & $15.24\pm0.02$ & $14.48\pm0.02$ & $13.56\pm0.03$ \\

\enddata
\end{deluxetable}
\clearpage

\begin{deluxetable}{cccc} \footnotesize  \tablecaption{Observation 
Intervals and the Dates \label{tab:int}} 
\tablehead{ \colhead{Interval \#} & \colhead{MJD} & \colhead{Periodic} & 
\colhead{Decay (e-folding timescale)} } 
\startdata
1 & 51180-51229 & No & 124 d \\
2 & 51230-51313 & No & 27 d\tablenotemark{a} \\
3 & 51314-51345 & Yes & 22.6 d\tablenotemark{b} \\
4 & 51346-51420 & Yes & 362 d\tablenotemark{c} \\
\enddata
\tablenotetext{a}{The decay corresponds to the light curve between 51244 and 
51300 only}
\tablenotetext{b,c}{The decay is superposed with a sinusoidal component}

\end{deluxetable}
\clearpage

\begin{figure}
\plotone{fig1.ps}
\figurenum{1}
\caption{}
\end{figure}

\begin{figure}
\plotone{fig2.ps}
\figurenum{2}
\caption{}

\end{figure}

\begin{figure}
\plotone{fig3.ps}
\figurenum{3}
\caption{}
\end{figure}

\begin{figure}
\plotone{fig4.ps}
\figurenum{4}
\caption{}
\end{figure}

\begin{figure}
\plotone{fig5.ps}
\figurenum{5}
\caption{}
\end{figure}

\begin{figure}
\plotone{fig6.ps}
\figurenum{6}
\caption{}
\end{figure}

\begin{figure}
\plotone{fig7.ps}
\figurenum{7}
\caption{}
\end{figure}

\begin{figure}
\plotone{fig8.ps}
\figurenum{8}
\caption{}
\end{figure}

\begin{figure}
\plotone{fig9.ps}
\figurenum{9}
\caption{}
\end{figure}

\begin{figure}
\plotone{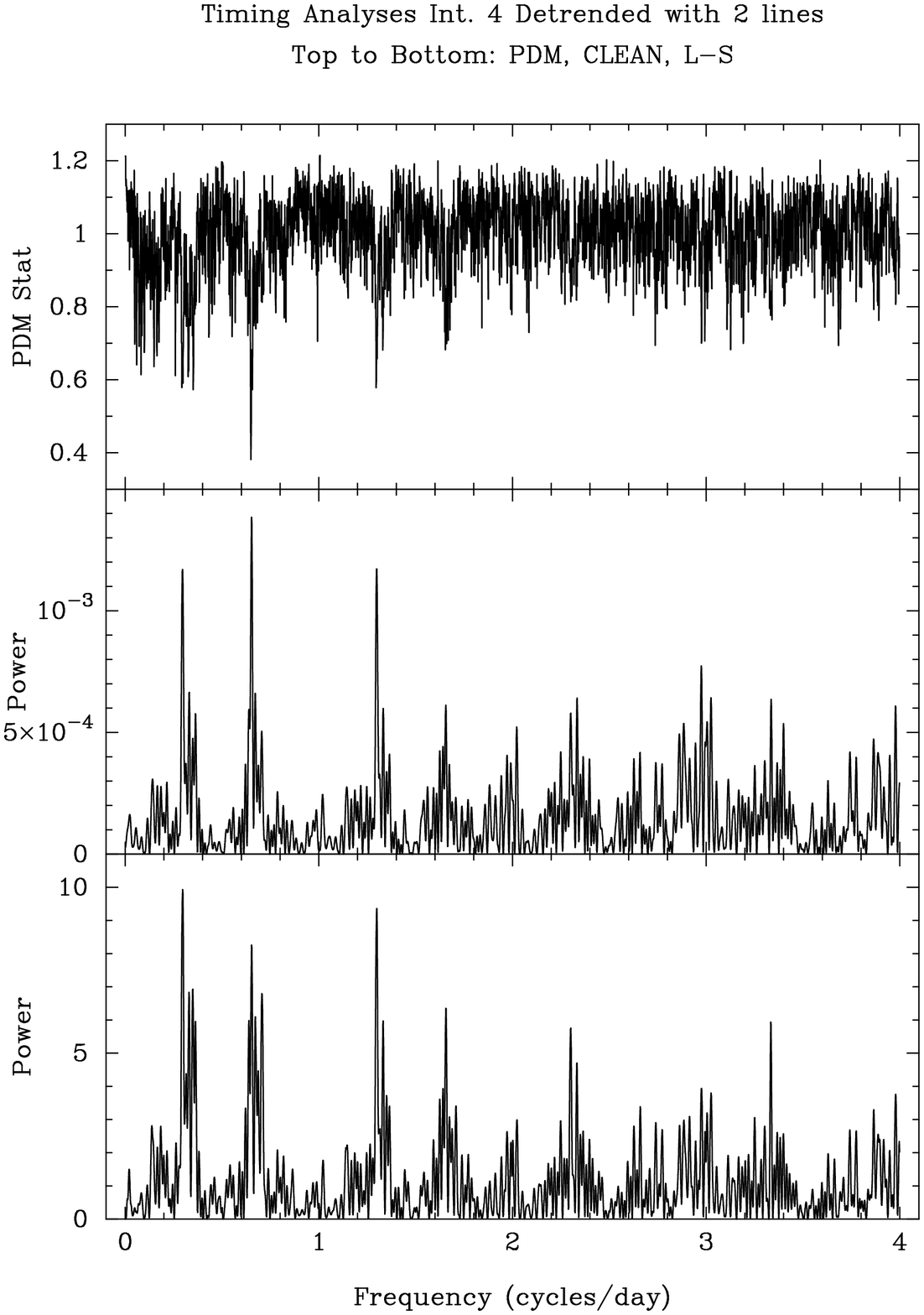}
\figurenum{10}
\caption{}
\end{figure}

\begin{figure}
\plotone{fig11.ps}
\figurenum{11}
\caption{}
\end{figure}

\begin{figure}
\plotone{fig12.ps}
\figurenum{12}
\caption{}
\end{figure}

\begin{figure}
\plotone{fig13.ps}
\figurenum{13}
\caption{}
\end{figure}

\begin{figure}
\plotone{fig14.ps}
\figurenum{14}
\caption{}
\end{figure}

\begin{figure}
\plotone{fig15.ps}
\figurenum{15}
\caption{}
\end{figure}

\end{document}